\input{aipcheck}

\documentclass[
    ,final            
  ]
  {aipproc}

\layoutstyle{6x9}

\begin{document}

\title{Electroweak scale neutrinos and Higgses}


\author{Alfredo Aranda}{
address={Facultad de Ciencias, CUICBAS, Universidad de Colima, \\
Bernal D\'iaz del Castillo No. 340, Colima, Colima, M\'exico.}, 
altaddress={Dual CP Institute of High Energy Physics (DCP-08-05).}
}

\begin{abstract}
We present two different models with electroweak scale right-handed neutrinos.
One of the models is created under the constraint that 
any addition to the Standard Model must not introduce new higher scales. 
The model contains right-handed neutrinos with electroweak scale masses and
a lepton number violating singlet scalar field. The scalar phenomenology
is also presented. The second model is a triplet Higgs model where again
the right-handed neutrinos have electroweak scale masses. In this case the
model has a rich scalar phenomenology and in particular we present the analysis
involving the doubly charged Higgs.
\end{abstract}

\maketitle


\section{Introduction}
We present two recent models~\cite{Aranda:2007dq,Aranda:2008ab}
of electroweak scale right-handed neutrinos and their scalar phenomenology.
First we describe a model based on the idea that given
our current (experimental) knowledge of particle physics one should explore 
a "truly minimal" extension of the Standard Model (SM). We
consider the possibility of having just one scale associated with
all the high energy physics (HEP) phenomena. Thus we propose
a minimal extension of the SM where new phenomena associated to
neutrino physics can also be explained by physics at the Electroweak
(EW). We then review a recent model~\cite{Hung:2006ap} in which the RH neutrinos that
participate in the seesaw mechanism are {\em active} in the sense that they
are {\em electroweak nonsinglets}. If they are not too heavy, they can be produced
at colliders and the seesaw mechanism could be tested. The right-handed neutrinos 
of~\cite{Hung:2006ap} are members of SM doublets of mirror leptons and
their Majorana masses are linked to EW scale through a coupling with a Higgs
triplet that develops an EW scale VEV. In this model, the sources of the SM SSB
include Higgs triplets.

\section{Minimal model}
Based on the minimalistic constraint described above we assume
\begin{itemize}
    \item SM particle content and gauge interactions.
    \item Existence of three RH neutrinos with a mass scale of
    EW size.
    \item Global U(1)$_L$ spontaneously (and/or explicitly) broken
    at the EW scale by a single complex scalar field.
    \item All mass scales come from spontaneous symmetry breaking
    (SSB). This leads to a Higgs sector that includes a Higgs
    SU(2)$_L$ doublet field $\Phi$ with hypercharge $1$
    (i.e. the usual SM Higgs doublet) and a SM singlet complex
    scalar field $\eta$ with lepton number $-2$.
\end{itemize}

The terms of the Lagrangian relevant for Higgs and neutrino physics are
${\cal L}_{\nu H}={\cal L}_{\nu y} - V$, with
\begin{eqnarray} \label{yukawas}
    {\cal L}_{\nu y}= -y_{\alpha i}\bar{L}_{\alpha}N_{Ri}\Phi
    -\frac{1}{2}Z_{ij}\eta \bar{N}_{Ri}^cN_{Rj} + h.c. \ ,
\end{eqnarray}
where $N_R$ represents the RH neutrinos, $\psi^c=C\gamma^0\psi^*$
and $\psi_R^c\equiv(\psi_R)^c=P_L\psi^c$ has left-handed chirality.
The scalar potential is given by
\begin{eqnarray}
\label{potential1}\nonumber
    V & = &
    \mu_D^2\Phi^{\dagger}\Phi+\frac{\lambda}{2}\left(\Phi^{\dagger}\Phi\right)^2
    + \mu_S^2\eta^*\eta +
    \lambda^{\prime}\left(\eta^*\eta\right)^2  \\
    & + & \kappa \left(\eta\Phi^{\dagger}\Phi + h.c. \right)
    +\lambda_m \left(\Phi^{\dagger}\Phi\right)\left(\eta^*\eta\right) \ .
\end{eqnarray}
Note that the fifth term in the potential breaks explicitly the
U(1) associated to lepton number.

It is useful to define the scalar
mass eigenstates through\begin{eqnarray} \label{scalarfield}
     {\cal H} = \left( \begin{array}{c}
     \phi^0 \\ \rho \end{array} \right) =
     \left( \begin{array}{cc}
     \cos\alpha & -\sin\alpha \\ \sin\alpha & \cos\alpha \end{array} \right)
     \left( \begin{array}{c}
     h \\ H \end{array} \right) \ ,
\end{eqnarray}
where we have used the following relations:
\begin{eqnarray} \label{vevs}
    \Phi = \left( \begin{array}{c} 0 \\ \frac{\phi^0 + v}{\sqrt{2}}
    \end{array} \right) \ \ {\rm and} \ \ \eta = \frac{\rho + u + i
    \sigma}{\sqrt{2}} \ .
\end{eqnarray}
Using these definitions the Lagrangian becomes
\begin{eqnarray}
     \label{yukawas2} \nonumber
     {\cal L}_{\nu y} & \supset & -y_{\alpha i} \bar{\nu}_{L\alpha}N_{Ri}\frac{\phi^0}{\sqrt{2}}
     -\frac{1}{2} Z_{ij}    \frac{(\rho+i\sigma)}{\sqrt{2}} \bar{N}_{Ri}^c N_{Rj} + h.c. \\ \nonumber
     & = & \left( -\frac{y_{\alpha i}}{\sqrt{2}}\bar{\nu}_{L\alpha}N_{Ri}
     (c_{\alpha} \ h -s_{\alpha} \ H) + h.c. \right) - \left(\frac{i}{2\sqrt{2}}
     Z_{ij} \bar{N}_{Ri}^c N_{Rj} \sigma + h.c. \right) \\
     & - & \left( \frac{1}{2\sqrt{2}} Z_{ij} \bar{N}_{Ri}^c N_{Rj} (s_{\alpha} \ h +
     c_{\alpha} \ H) + h.c \right) \ .
\end{eqnarray}

We are interested in EW scale RH neutrinos. The Dirac part on the other hand 
will be constrained from the seesaw. Writing the neutrino mass matrix as
\begin{eqnarray} \label{mneutrino} m_{\nu} = \left( \begin{array}{cc} 0 & m_D \\
m_D & M_M \end{array}\right) \ , \end{eqnarray} where $(m_D)_{\alpha i} =
y_{\alpha i}v/\sqrt{2}$.  As an example lets consider the third
family of SM fields and one RH neutrino, thus Eq.({\ref{mneutrino})
becomes a $2\times 2$ matrix. Assuming $m_D << M_M$ we obtain the
eigenvalues $m_1=-m_D^2/M_M$ and $m_2=M_M$ and by requiring $m_1
\sim$~O(eV) and $m_2 \sim (10 -100)$~GeV and using $v=246$~GeV we
obtain an upper bound estimate for the coupling $y_{\tau i} \leq
10^{-6}$.

The mass eigenstates are denoted by $\nu_1$ and $\nu_2$ and are such that
\begin{eqnarray} \label{transformation} \nonumber
\nu_{\tau} & = & \cos\theta \ \nu_{L1} + \sin\theta \ \nu_{R2} \\
N & = & -\sin\theta \ \nu_{L1} +\cos\theta \ \nu_{R2} \ , \end{eqnarray} where
$\theta=\sqrt{m_D/m_2} \approx 10^{-(5 - 6)}$.

The relevant terms in the Lagrangian become \begin{eqnarray}
\label{relevantterms} \nonumber {\cal L} & \supset & \left[ h
\bar{\nu}_{L1}^c \nu_{L1} \left( -\frac{Z}{2\sqrt{2}}s_\theta^2
s_\alpha \right)+  h \bar{\nu}_{R2}^c \nu_{R2}
\left(- \frac{Z}{2\sqrt{2}} c_\theta^2 s_\alpha \right) + h.c. \right] \\
& + & h \bar{\nu}_{L1} \nu_{R2}
\left(\frac{y_\nu}{\sqrt{2}}(s_{\theta}^2-c_{\theta}^2)c_\alpha
\right) +  h \bar{\nu}_{R2} \nu_{L1}
\left(\frac{y_\nu}{\sqrt{2}}(s_\theta^2-c_{\theta}^2) c_\alpha
\right) \ , \end{eqnarray} where $y_{\nu}^*=y_{\nu}$ and $Z \equiv Z_{11}$.

In this work we are interested in presenting the results for the
Higgs decays to neutrinos and their signatures in this model. 
Using Eq.~(\ref{relevantterms}) we compute the following decay
widths~\footnote{All SM decay widths will have an extra factor of
$c_{\alpha}^2$}: \begin{eqnarray} \label{widths}
\Gamma(h\rightarrow \bar{\nu}_1\nu_1) & = & \frac{m_h}{64\pi}|Z|^2 s_{\theta}^4 s_{\alpha}^2 \ , \\
\Gamma(h\rightarrow \bar{\nu}_2\nu_2) & = & \frac{m_h}{64\pi}|Z|^2 c_{\theta}^4 s_{\alpha}^2
\left(1-\frac{4m_2^2}{m_h^2}\right)^{3/2} \ , \\
\Gamma(h\rightarrow \bar{\nu}_1\nu_2) & = &
\frac{m_h}{16\pi}y_{\nu}^2(s_{\theta}^2-c_{\theta}^2)^2 c_{\alpha}^2
\left(1-\frac{m_2^2}{m_h^2}\right)^2 \ . \end{eqnarray}

We have computed the branching ratios for the Higgs decays and the
results are presented in Figure~\ref{fig:branching}. In each plot we
have included the results for three values of $\cos\alpha$ ($0.1$,
$0.5$ and $0.9$). The two graphs correspond to the values of $m_2
= 60$ and $100$~GeV respectively. Only the dominant
contributions are shown for clarity, i.e. $h \to \nu_2\bar{\nu_2}, \
b\bar{b} \ \rm{and} \ \tau \bar{\tau}$. It is interesting to note
that for the whole range where it is possible, the decay $h \to
\nu_2 \bar{\nu_2}$ dominates in all three cases. This is a clear
distinctive signature of our model.
In order to study the specific signatures that would be observed in
this scenario, we consider the $\nu_2$ decays. In
Table~\ref{tab:signatures} we present the possible signatures of
these decays.
\begin{table}
    \begin{tabular}{cccc}
      \hline
      {\rm Higgs decay} & $\nu_2 \rightarrow \nu_1 Z^*$ &
      $\nu_2 \rightarrow l W^*$ &  $\nu_2 \rightarrow \nu_1 \gamma$ \\
      \hline
      $h\rightarrow \nu_1\nu_2$ & $l^+l^- + {\rm inv.}$ & $l+l^{\prime}+ {\rm inv.}$ &
      $\gamma + {\rm inv.}$ \\
      & $q\bar{q} + {\rm inv.}$ & $l+q\bar{q}^{\prime}  + {\rm inv.}$ & \\
      $h\rightarrow \nu_2\nu_2$ & $l^+l^- + l^+l^- + {\rm inv.}$ &
      $l + l^{\prime} + l^{\prime\prime}+ l^{\prime\prime\prime} + {\rm inv.}$ & \\
      & $l^+l^- + q\bar{q} + {\rm inv.}$ & $l + l^{\prime} + l^{\prime\prime}+ q\bar{q}  + {\rm inv.}$ &
      $\gamma + \gamma  + {\rm inv.}$ \\
      & $q\bar{q} + q\bar{q}  + {\rm inv.}$ & $l+l^{\prime}+ q\bar{q}+q\bar{q} + {\rm inv.}$ & \\
      $h\rightarrow \nu_1\nu_1$ & - & - & - \\
      \hline
    \end{tabular}
    \caption{Signatures for the Higgs decays considered in the text.}
    \label{tab:signatures}
\end{table}

Since we are interested in a Higgs mass in the natural window of
$100-200$~GeV, and in neutrino masses such that they can appear in
Higgs decays, we will consider neutrino masses of order
$10-100$~GeV, therefore we need to consider the 3-body decays $\nu_2
\to \nu_1 + V^* (\to f \bar{f}')$, where $V^*= W^* , Z^*$:
\begin{eqnarray} \label{widthN}
\Gamma=\frac{m_2^5}{256\pi^3}\frac{5}{16}\frac{(B^2+C^2)(a_f^2+b_f^2)}{M_V^4}
\ , \end{eqnarray} where 
\begin{eqnarray} \label{parameters} \nonumber (V = W) &
\rightarrow & \left\{ \begin{array}{c}
a_f  =  -b_f \equiv a = \frac{g}{2\sqrt{2}} \\
B = -C = a \ s_{\theta} \end{array} \right. \ \ \ \ 
(V = Z)  \rightarrow  \left\{
\begin{array}{c}
a_f = \frac{g}{2c_w}(T^3_f-2Q_f s_w^2) \\ \nonumber
b_f = -\frac{g}{2c_w} T^3_f \\
B = a_{\nu} \ c_{\theta}s_{\theta} \\
C = b_{\nu} \  c_{\theta}s_{\theta} \end{array} \right. \end{eqnarray}

The branching ratios for these processes are presented in table~\ref{tab:nu2decays}. 
We show the results for $m_2 = 100$~GeV as the results are similar in all the $m_2$ range
considered in this paper. We find that the dominant contributions
are the ones associated to the $W^*$ decay process.
\begin{table}[ht]
        \begin{tabular}{c|cccccc}
            \hline
            $m_2 (GeV)$ & $\nu \ l^+ \ l^-$ & $\nu \ \nu \ \nu$ & $\nu
            \ q_u \ \bar{q}_u$ & $\nu \ q_d \ \bar{q}_d$ & $l^{\pm} \ l^{\pm} \ \nu$ &
            $l^{\pm} \ q \ \bar{q}^{\prime}$ \\
            \hline
            $100$ & $0.008$& $0.015$ & $0.018$ & $0.034$ & $0.308$ &
            $0.617$ \\
            \hline
        \end{tabular}
        \caption{Branching ratios for the $\nu_2$ three body decays discussed in the text. The results
        correspond to $m_2=100$~GeV and do not depend strongly on the value of $m_2$.}
        \label{tab:nu2decays}
\end{table}

\section{Model with Higgs triplets}
We now review the basic structure of the second model.
The full description of the scalar sector involving the triplet fields can be found
in~\cite{Chanowitz:1985ug,Accomando:2006ga,Gunion:1989ci,Han:2007bk},
here we briefly review the extension of the basic model to include electroweak
neutrinos.

In addition to the SM particle content the model of~\cite{Hung:2006ap} contains the
additional fields shown in table~\ref{tab:fields}. There is also an additional global U(1)$_M$ 
symmetry under which \begin{eqnarray} \label{u1m} L_R^M, \ e_L^M \rightarrow e^{i\theta_M}L_R^M, \
e_L^M; \ \ \tilde{\chi} \rightarrow e^{-2i\theta_M}\tilde{\chi}, \ \
\phi_S \rightarrow e^{-i\theta_M}\phi_S \ , \end{eqnarray} and all other
fields are singlets. This global symmetry was invoked in order to
avoid certain terms as indicated below and was explained in detail in~\cite{Hung:2006ap}. 
\begin{table}[ht] 
  \begin{tabular}{|c|c|c|} \hline
    {\rm Additional \ fields} & SU(2)$_W$ & U(1)$_Y$ \\ \hline \hline
    $L_R^M =\left( \nu_R \ \ e_R^M \right)$ & ${\bf 2}$ & $0$ \\ \hline
    $\tilde{\chi}=\left(\chi^0 \ \ \chi^+ \ \ \chi^{++} \right)^T$ & ${\bf 3}$ & $-2$ \\ \hline
    $\xi=\left(\xi^+ \ \ \xi^0 \ \ \xi^{+} \right)^T$ & ${\bf 3}$ & $0$ \\ \hline
    $e_L^M$ \ \& \ $\phi_S$  & ${\bf 1}$ & $0$ \\ \hline
  \end{tabular}
  \caption{\label{tab:fields} Additional field content}
\end{table}

Since $\nu_R$ is not an SU(2)$_L$ singlet, it does not
couple to $\bar{L}_L\tilde{\Phi}$. Instead, the Dirac neutrino mass
comes from the term 
${\cal L}_S=-g_{sl}\bar{L}_L\phi_SL_R^M + h.c.$, which leads to
$M_{\nu}^D=g_{sl}v_s$, where $\langle\phi_S\rangle=v_S$ and thus the
neutrino Dirac mass is independent of the EW scale.

RH neutrinos must have a mass $> M_Z/2$ in order not to
contribute to the $Z$ width. This is accomplished with the $Y=-2$
triplet $\tilde{\chi}$ through the term
$g_ML_R^{M,T}\sigma_2\tau_2\tilde{\chi}L_R^M$,
which leads to $M_R=g_M v_M$, with $\langle\chi^0\rangle=v_M$ and where $v_M = O(\Lambda_{EW})$. This allows to have
EW-scale masses for the right-handed neutrinos without having to
fine-tune the Yukawa coupling $g_M$ to be abnormally small.

The U(1)$_M$ symmetry is introduced in order to forbid the terms
$g_LL_L^T\sigma_2\tau_2\tilde{\chi}L_L$ and
$L_L^T\sigma_2\tau_2 \tilde{\chi}L_R^M$ at tree level. 
The main consequence of this is that
the Dirac mass for the neutrinos comes from $v_s$ exclusively and
the Majorana mass, $M_L$, for the left-handed neutrinos arises at the one-loop level
and can be much smaller than $M_R$.

Taking all of this into consideration one obtains the following
Majorana mass matrix: 
\begin{eqnarray} \label{MM} {\cal M}=\left(
\begin{array}{cc}M_L & m_{\nu}^D \\ m_{\nu}^D & M_R \end{array}
\right) \ , 
\end{eqnarray}
where  $M_L \sim \epsilon (m_{\nu}^D)^2/M_R< 10^{-2} (m_{\nu}^D)^2/M_R$. 

We are interested in the scenario where $g_{sl} \sim$~O($g_M$) and
$v_M >> v_S$. In this case, the eigenvalues of ${\cal M}$ become
$-(g_{sl}^2/g_M)(v_s/v_m)v_s(1-\epsilon)$ and $M_R$, where $\epsilon
< 10^{-2}$. Now, since $v_M \sim \Lambda_{EW}$, and using the bound
$m_{\nu}\leq 1$~eV, we have $ v_S \approx \sqrt{(1
{\rm eV}) \times v_M}\sim {\rm O}(10^{5-6}{\rm eV})$. 

The kinetic part of the Higgs Lagrangian is given by 
\begin{eqnarray} \label{Lhiggs}
  {\cal L}_{kin}=\frac{1}{2}Tr[(D_{\mu}\Phi)^{\dagger}(D^{\mu}\Phi)]+
  \frac{1}{2}Tr[(D_{\mu}\chi)^{\dagger}(D^{\mu}\chi)] +
  |\partial_{\mu}\phi_s|^2\ .
\end{eqnarray} 
The potential (for $\Phi$ and $\chi$)\footnote{We work under the 
assumption that $\phi_S$ does not couple with 
the other Higgses at tree level. We choose to work with this assumption because
the coupling generated at loop level, through the $\phi_S$ couplings to 
SM left-handed fermions and to mirror right-handed fermions, can be very 
small~\cite{Hung:2007ez}} to be considered
is~\cite{Chanowitz:1985ug} 
\begin{eqnarray} \label{potential} \nonumber
  V(\Phi,\chi)&=&\lambda_1(Tr\Phi^{\dagger}\Phi-v_2^2)^2+
  \lambda_2(Tr\chi^{\dagger}\chi-3v_m^2)^2 \\ \nonumber &+&
  \lambda_3(Tr\Phi^{\dagger}\Phi - v_2^2+Tr\chi^{\dagger}\chi -
  3v_m^2)^2 \\ \nonumber &+& \lambda_4 ( Tr\Phi^{\dagger}\Phi
  Tr\chi^{\dagger}\chi - 2Tr\Phi^{\dagger} T^i\Phi T^j\cdot
  Tr\chi^{\dagger}T^i\chi T^j)
  \\ &+& \lambda_5[3Tr\chi^{\dagger}\chi \chi^{\dagger}\chi
    -(Tr\chi^{\dagger}\chi)^2] \ . 
\end{eqnarray}

Note that this potential is invariant under $\chi \rightarrow
-\chi$. When $\chi$ gets a vev $\langle\chi\rangle=diag(v_M,v_M,v_M)$ it
breaks the global symmetry SU(2)$_L \times$ SU(2)$_R$ down to the
custodial SU(2)$_C$. It was shown in ~\cite{Chanowitz:1985ug, georgi}
that the structure of the VEV is dictated by the proper vacuum alignment.
Now, using $\langle\Phi\rangle=v_2/\sqrt{2}$,
the $W$ and $Z$ masses can be obtained from Eq.~(\ref{Lhiggs}) and
are given by $M_W=gv/2$ and $M_Z=M_W/\cos\theta_W$, with $v^2=v_2^2+8v_M^2$,
with $v \approx 246\,{\rm GeV}$.  This gives rise to $\rho=1$ at tree level.

A convenient parametrization can be made by defining $\cos\theta_H =
c_H \equiv v_2/v$ and thus $\sin\theta_H=s_H\equiv 2\sqrt{2}v_M/v$.
Using these parameters we can see that $\tan\theta_H=t_H$
characterizes the amount of the $W$ mass coming from either the
doublet or the triplet scalars.

If the potential preserves the SU(2)$_C$ then the fields get arranged 
in the following manner (based on their transformation properties under 
the custodial SU(2)): 
\begin{eqnarray} \label{scalarfields} 
  {\rm five-plet} &\rightarrow& H_5^{\pm
    \pm}, \ H_5^{\pm}, \ H_5^0 \leftrightarrow \ {\rm degenerate} \\
  {\rm three-plet} &\rightarrow& H_3^{\pm}, \ H_3^0 \leftrightarrow \
  {\rm degenerate} \\ 2 - {\rm singlets} &\rightarrow& H_1^0, \
  H_1^{0\prime} \leftrightarrow \ {\rm Only \ these \ can \ mix} \ ,
\end{eqnarray} 
where the definitions and Feynman rules for vector boson couplings can be found
in~\cite{Gunion:1989ci}. In the search for the Higgs scalars discussed in this work, it
is important to know what those scalars couple to. The couplings
of this extended Higgs sector can be found in ~\cite{georgi} while
the Feynman rules for scalar fermion couplings including the mirror fermions
are presented in~\cite{Aranda:2008ab}.

In this paper we present the results obtained for the doubly charged Higgs phenomenology. 
The complete numerical analysis of this model can be found in~\cite{Aranda:2008ab}.

The presence of a doubly charged Higgs in this model provides with interesting
phenomenology. Furthermore, the phenomenology of this model is specific and
different from that of the general two triplets model due to the following
observations:

\begin{itemize}
\item Due to the U(1)$_M$ symmetry of the model or its embedding
in a Pati-Salam type of quark-lepton unification, the term proportional to
$l_l^T\sigma_2\tau_2\tilde{\chi}l_L$ is not allowed and thus the decay
$\Gamma(\chi^{++}\to l^+l^+)$ is not present.
\item The presence of mirror fermions and $\phi_S$ allows for the decays $\Gamma(\chi^{++}
\to l_i^M \ l_j^M)$ and $\Gamma(\chi^{++}\to l \ \phi_S \ l_M)$ or
even $\Gamma(\chi^{++}\to l l \phi_S \phi_S)$.
\end{itemize}

Using the expressions for the $\chi^{++}$ decays in~\cite{Aranda:2008ab} we can compute 
the branching ratios. In the following analysis we have made the following assumptions:

\begin{itemize}
\item $g_M$ and $g_{sl}$ are proportional to the identity matrix and so, in each of the
expressions above, $g_M$ and $g_{sl}$ represent numbers.
\item The model requires $g_{sl}^2/g_M \sim$~O(1). We have chosen numbers of O(1) for
both couplings and for the numerical results presented below they have been set to
$g_M = 0.7$ and $g_{sl} = 0.8$.
\end{itemize}

Given these assumptions we compute the following branching ratios: 
$B(\chi^{++}\to l^+_M l^+_M)$, $B(\chi^{++}\to W^+ W^+)$, $B(\chi^{++}\to H^+_3 W^+)$,
$B(\chi^{++}\to l^+ \nu W^+)$ and $B(\chi^{++}\to l^+ \phi_S l_M^+)$. 

Figure~\ref{fig:br1} shows the branching ratios for three different values
of $\sin\theta_H$ and for small values of the mirror fermions masses (taken
to be degenerate) $m_{lM} = 50\, {\rm GeV}$. We can see that the dominant one always
corresponds to $B(\chi^{++}\to l_Ml_M)$, while the relative dominance of
the other channels depends on $\sin\theta_H$.

Similar results are obtained for larger $m_{lM}$ as can be seen in figure~\ref{fig:br2}
where we show the branching ratios for $m_{lM}=100\, {\rm GeV}$.

\begin{theacknowledgments}
We thank the organizers of the XIII Mexican School of Particles and Fields
and would like to acknowledge our collaborators in this work: O. Blanno, J.L. D\'iaz-Cruz,
J. Hern\'andez-S\'anchez and P.Q. Hung. This work was partially supported by Conacyt.
\end{theacknowledgments}





\IfFileExists{\jobname.bbl}{}
 {\typeout{}
  \typeout{******************************************}
  \typeout{** Please run "bibtex \jobname" to optain}
  \typeout{** the bibliography and then re-run LaTeX}
  \typeout{** twice to fix the references!}
  \typeout{******************************************}
  \typeout{}
 }

\vspace{1cm}

\begin{figure}[ht]
    \includegraphics[width=7cm]{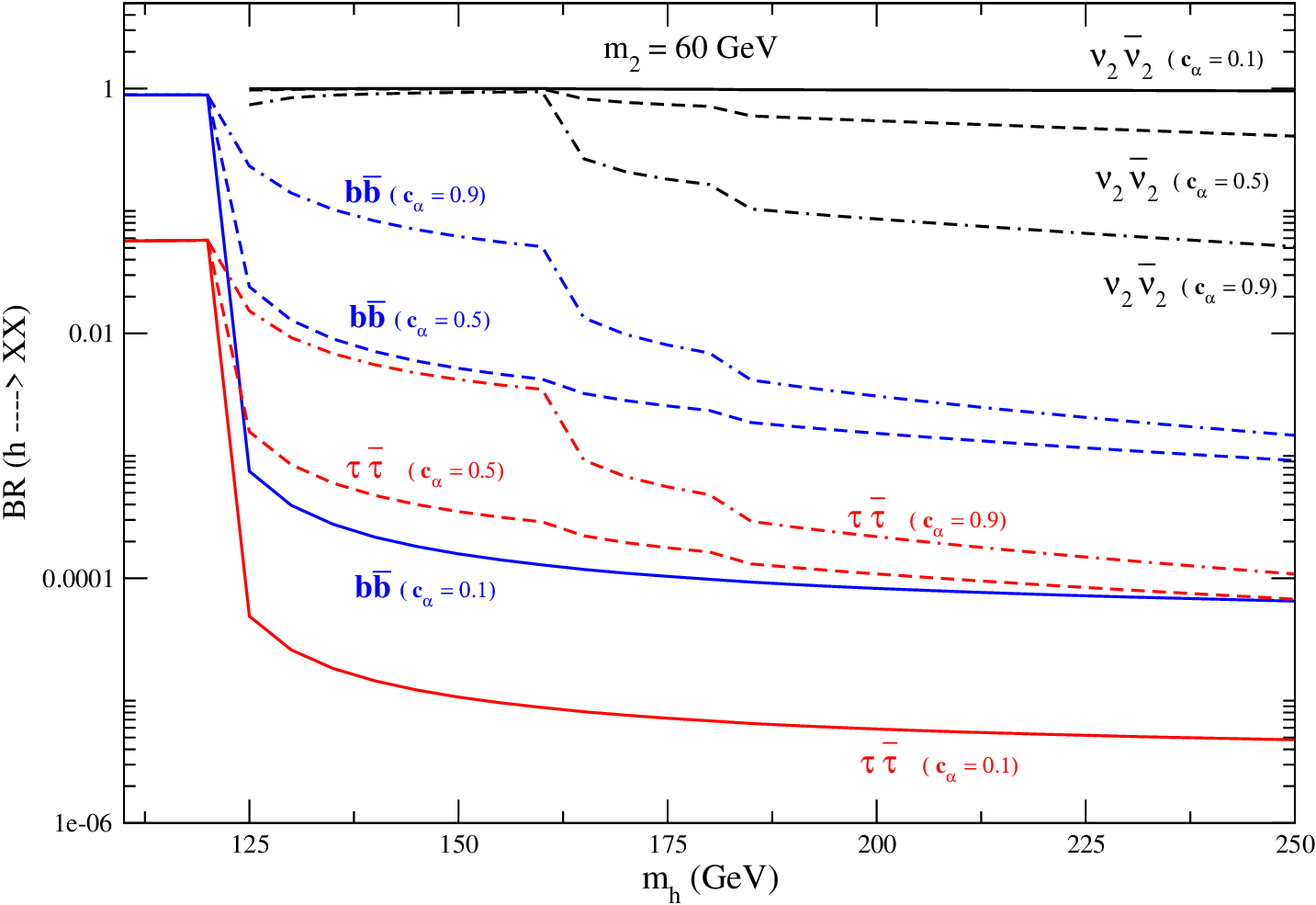}
    \includegraphics[width=7cm]{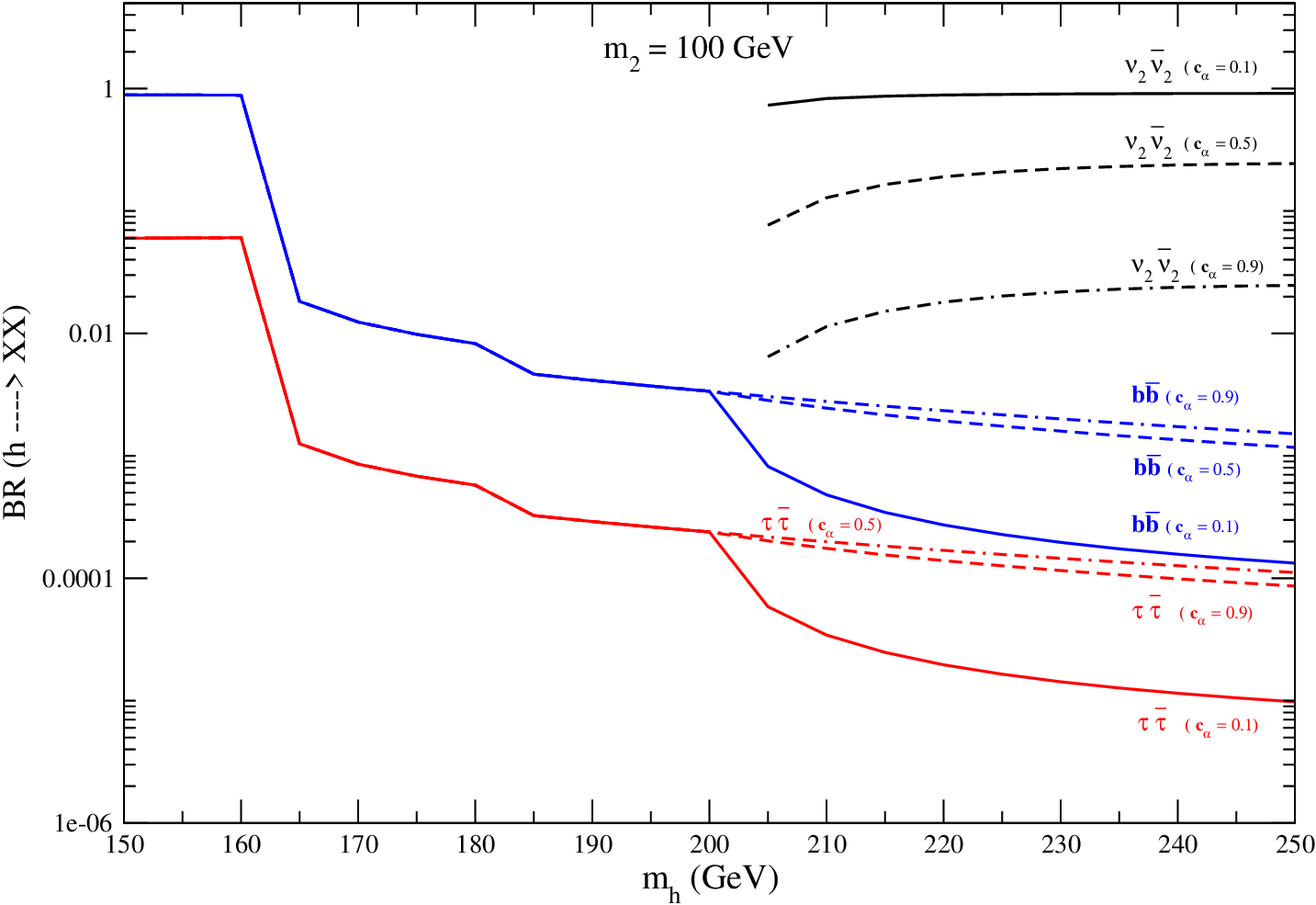}
    \caption{Dominant branching ratios for Higgs decays.
    Two cases are presented for $m_2 = 60 \ \rm{and} \ 100$~GeV
    respectively. Each plot includes results for the three values of
    $\cos\theta = 0.1, \ 0.5 \ \rm{and} \ 0.9$ as discussed in the text.}
    \label{fig:branching}
\end{figure}

\begin{center}
  \begin{figure}[ht]
    \includegraphics[width=5cm]{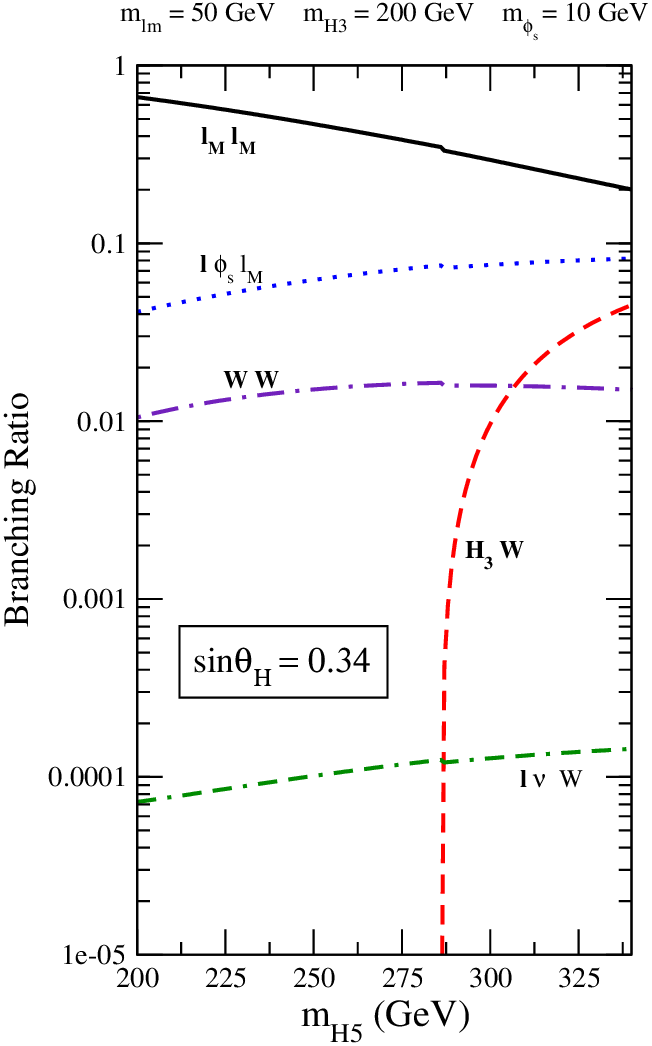}
    \includegraphics[width=5cm]{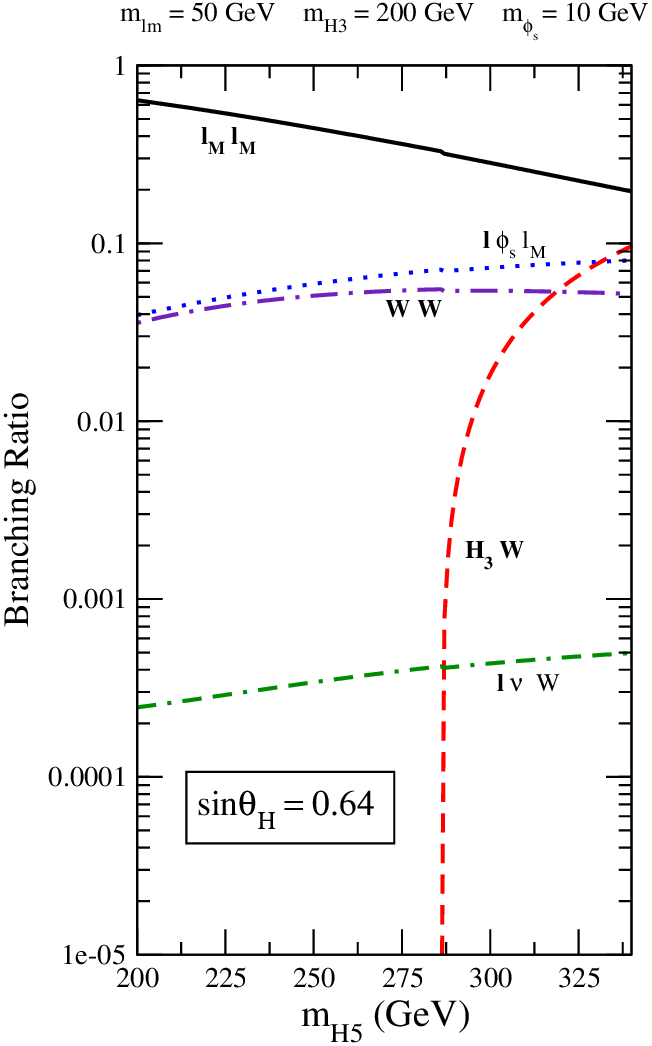}
    \includegraphics[width=5cm]{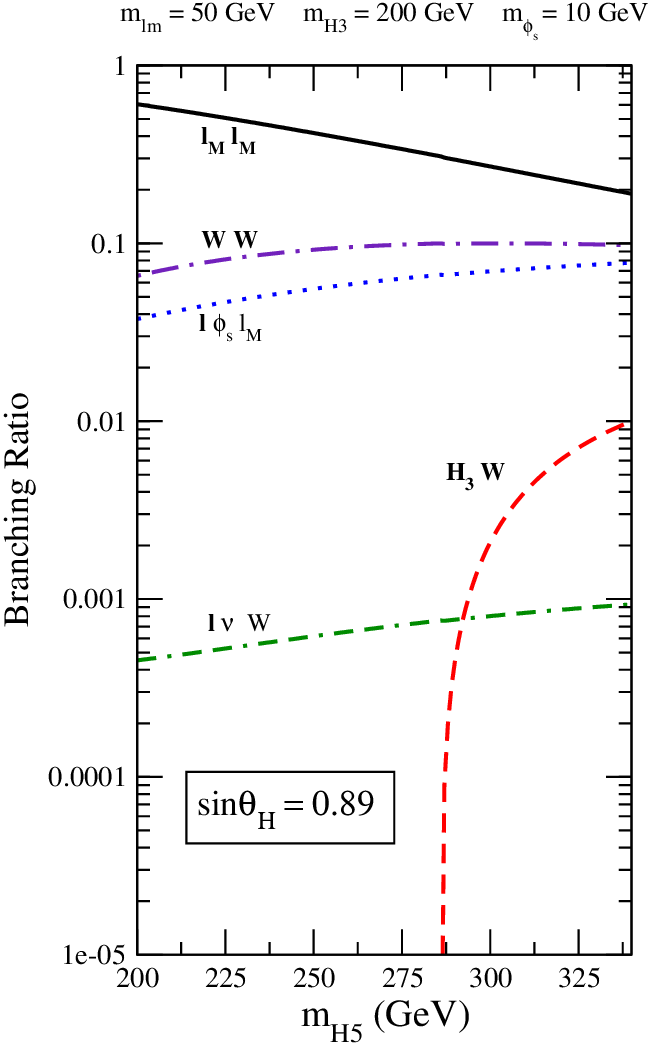}
    \caption{Branching ratios for $\chi^{++}$ as a function of its mass,
      for three different values of $\sin\theta_H$, and for a small $m_{lM}$.}
    \label{fig:br1}
  \end{figure}
\end{center}

\begin{center}
  \begin{figure}[ht]
    \includegraphics[width=5cm]{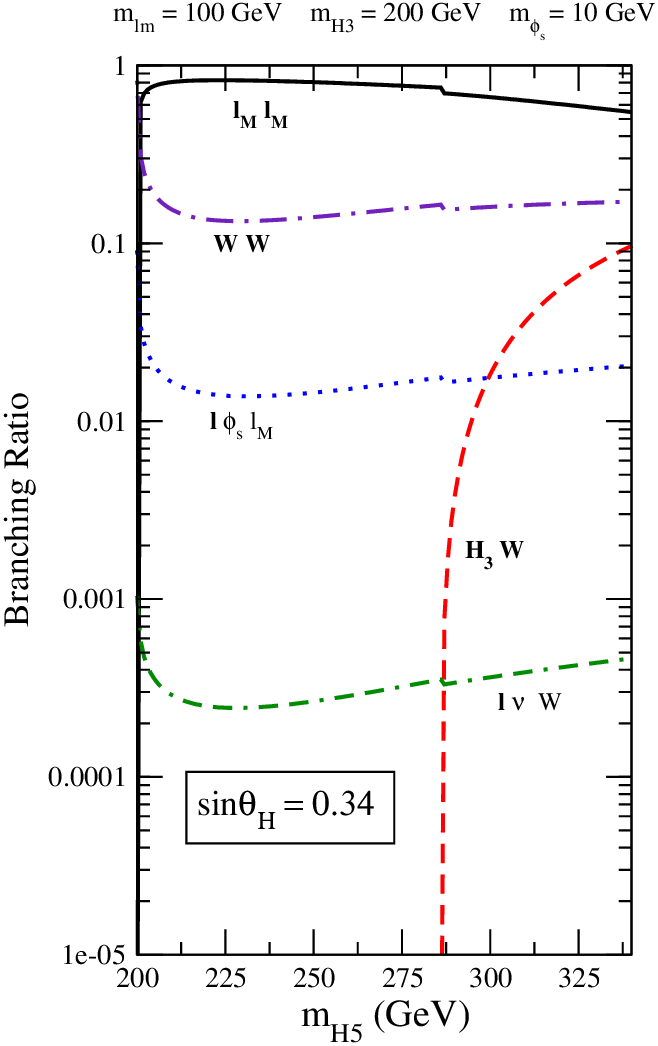}
    \includegraphics[width=5cm]{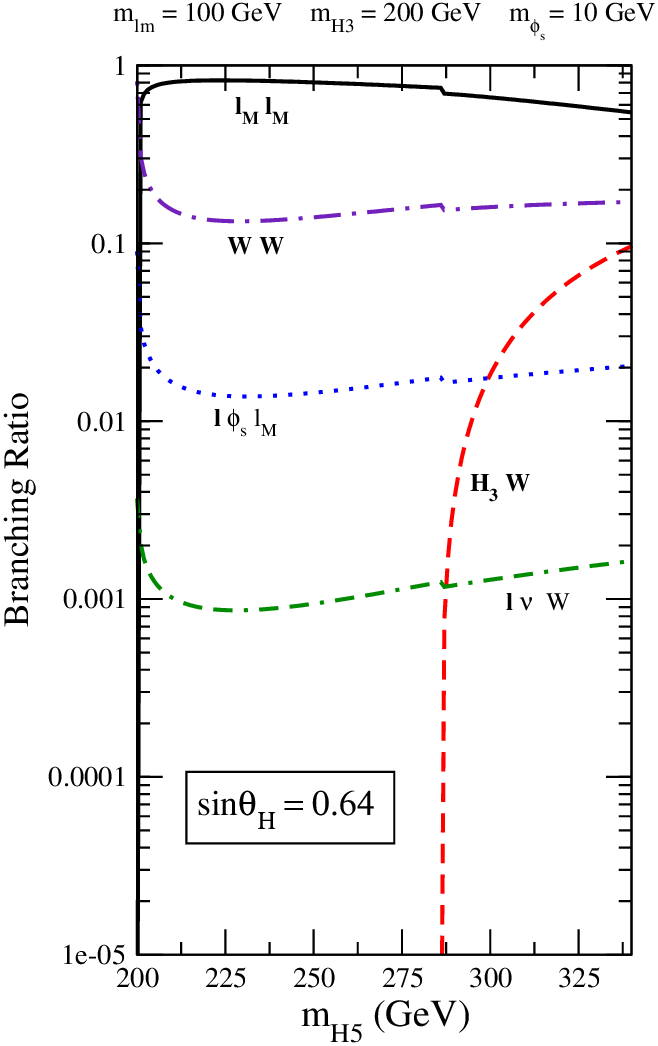}
    \includegraphics[width=5cm]{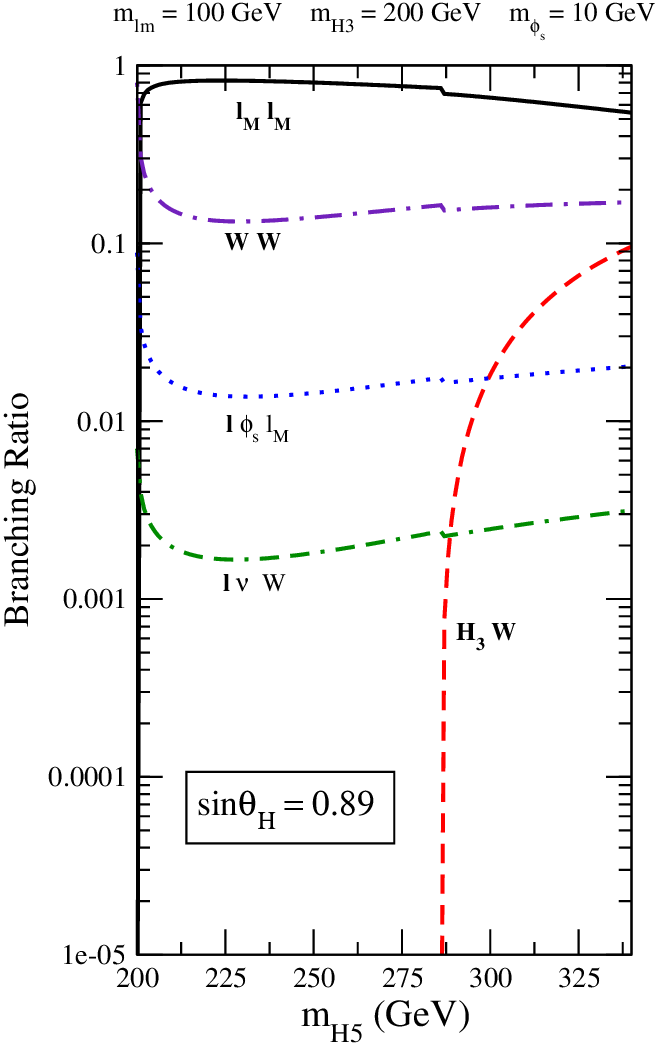}
    \caption{Same as before but with a heavier $m_{lM}$.}
    \label{fig:br2}
  \end{figure}
\end{center}

\end{document}